\documentclass[12pt, onecolumn]{IEEEtran}

\makeatletter
\def\ifundefined{\@ifundefined}
\makeatother

\usepackage{setspace}
\usepackage{algcompatible}
\usepackage{algorithmicx}
\usepackage{algpseudocode}
\usepackage{algorithm}
\usepackage{srcltx}
\usepackage{amsmath}
\usepackage{amssymb}
\usepackage{graphics}
\usepackage{graphicx}
\usepackage{cite}
\usepackage{psfrag}
\usepackage{epsfig}
\usepackage{subfigure}
\usepackage{amsthm}
\usepackage{thmtools}
\newcommand*{\qedb}{\hfill\ensuremath{\blacksquare}}

\newtheorem{thm}{Theorem}
\newtheorem{cor}{Corollary}

\newtheorem{remark}{Remark}

\newtheorem{lem}{Lemma}

\doublespace

\begin{document}

\renewcommand{\textfraction}{0}

\title{Cascading Failures in \\Finite-Size
Random Geometric Networks}

\author{Ali Eslami, ~\IEEEmembership{Member,~IEEE,} Chuan Huang, ~\IEEEmembership{Member,~IEEE,}\\ Junshan Zhang, ~\IEEEmembership{Fellow,~IEEE,} and Shuguang Cui, ~\IEEEmembership{Fellow,~IEEE}
\thanks{The material in this paper was presented in part at the 52nd Annual Allerton Conference on Communication, Control, and Computing, 2014.

A. Eslami and S. Cui are with the
Electrical and Computer Engineering Department, Texas A\&M University, College Station, TX, USA ~(email:\{eslami, cui\}@tamu.edu).

C. Huang and J. Zhang are with the
School of Electrical, Computer and Energy Engineering, Arizona State University, Tempe, AZ, USA ~(email: \{huangch, junshan.zhang\}@asu.edu).}}

\maketitle
\vspace{-.5 in}

\begin{abstract}
The problem of cascading failures in cyber-physical systems is drawing much attention in lieu of different network models for a diverse range of applications.
While many analytic results have been reported for the case of large networks, very few of them are readily applicable to finite-size networks. This paper studies cascading failures in finite-size geometric networks where the number of nodes is on the order of tens or hundreds as in many real-life networks. First, the impact of the \emph{tolerance parameter} on network resiliency is investigated.
We quantify the network reaction to initial disturbances of different sizes by measuring the damage imposed on the network. Lower and upper bounds on the number of failures are derived to characterize such damages. Such finite-size analysis reveals the decisiveness and criticality of taking action within the first few stages of failure propagation in preventing a cascade.
By studying the trend of the bounds as the number of nodes increases, we observe a phase transition phenomenon in terms of the tolerance parameter. The critical value of the tolerance parameter, known as the \emph{threshold}, is further derived.
The findings of this paper, in particular, shed light on how to choose the tolerance parameter appropriately such that a cascade of failures could be avoided.
\end{abstract}

\begin{keywords}
Cascading Failure, Finite-Size Complex Networks, Random Geometric Graph.
\end{keywords}

\section{introduction}
A cascading failure in a complex network is a phenomenon in which the failure of a small set of nodes triggers the failure of successive nodes, leading to the failure of a large fraction of the network eventually. There have been many types of cascading failure events that occurred in natural and man-made systems, from power grid and computer networks to political, economic, and ecological systems. Cascading failure is common in power grids, where a single failure of a fully loaded or slightly overloaded node (component) could set off more overloads, thereby taking down the entire system in a very short time. A few examples of power outages caused by cascading failures are the blackouts in northeast America in 2003, Italy in 2003, London in 2003, and  northern India in 2012.
Cascading failures can also occur in computer networks (such as the Internet), when a crucial router or node becomes overloaded. Network traffic then needs to be re-routed through an alternative path. This alternative path, as a result, may become overloaded, causing path break-down, and so on.

The problem of cascading failures in complex networks has been studied extensively \cite{Buldyrev10, Dobson05, Huang13, Dobson04, Kim10, Crucitti04, Rahnamay13, Lai13}, especially for large networks.
For the sake of tractability, different types of random graphs have been used to model complex networks, including Bernoulli random graphs, random geometric graphs, and scale-free graphs \cite{Watts02,Wang08,Asztalos14,Janson12,Kong10}.  Also, depending on the underlying applications, different models of failure propagation have been considered, where two popular categories of propagation rules are the \emph{degree-based} and \emph{load-based} propagation, respectively. In a degree-based propagation, the state of each node is determined by the states of all or part of its neighbors in the network \cite{Watts02,Janson12,Dobson04,Buldyrev10,Kong10}. For example, in \cite{Watts02}, each node is assigned a random threshold $\phi$, and it fails if at least a fraction $\phi$ of its neighbors fail.
On the other hand, in a load-based propagation, the state of a node is defined over the amount of load that it carries \cite{Wang08,Dobson05}. For instance in \cite{Wang08}, each node can carry a load up to its capacity, above which it becomes overloaded. An overloaded node fails and redistributes its load to its neighbors.

While the vast majority of the existing analytical studies are focused on large-scale networks, their findings can hardly be applied to the small or moderate size networks that we usually face in the real world. In this paper, we are concerned with providing rigorous analytical results for \emph{finite-size} networks. Furthermore, we are interested in studying cascading failures in networks  with geometric characteristics such as electrical power grids and wireless communication networks,  which could be well-modeled as \emph{random geometric graphs}. Indeed, random geometric graphs have been widely used in studying wireless networks (see \cite{Eslami10} and references therein). As expected, it is shown that geometry plays an important role in quantifying the topology of the smart grid communication and control networks \cite{Wang10}.

We adopt a \emph{load-based} failure propagation in this paper as it makes sense in a set of important applications such as the power grid and wireless networks. We assume that each node has a certain capacity,  part of which is used to carry a load  in normal conditions. If, for any reason, a node receives more load than its capacity, it fails and redistributes its load to its neighbors. A node here could be a component in a power grid, such as a transmission line or a regional transformer, which usually operates in normal conditions but is able to handle some higher loads up to a certain capacity. A node could also be a device in a wireless distributed storage network, or a routing hub in the Internet. In all these cases, a node could be assumed to operate under a certain load in normal situations, while it is able to handle a higher load up to a limit, if necessary.

The relative gap between the capacity and the normal load of a node is specified by the \emph{tolerance parameter} \cite{Wang08,Dobson05}. Tolerance parameter is a design parameter that plays an important role in network resiliency against a cascade. When resiliency is the priority, a larger tolerance parameter is desired as it enables the network to handle more severe operation disturbances. However, a larger tolerance parameter leads to a larger unused capacity that imposes higher costs. Therefore, it is crucial to obtain a clear understanding of the impact of tolerance parameter on network reactions to disruptions of different scales. In this paper, we characterize such reactions through analytical means in both finite and asymptotic regimes.

The rest of this paper is organized as follows. In the next section, we formally state the problem and explain the main contributions of this paper. In Section \ref{sec:prelim}, we provide some notation and preliminaries helpful for understanding the analysis. 
Sections \ref{sec:upper}, \ref{sec:asympupper}, and \ref{sec:lower} provide the main results and the bulk of the analysis. Finally, Section \ref{sec:conclusion} concludes the paper.

\section{Problem Formulation and Summary of Main Results}\label{sec:model}
In this section, we explain the models we use for the network, the initial disturbance, and the propagation of failures due to a disturbance. Furthermore, we introduce a method to quantify the overall damage caused by an initial disturbance. Given all that, we will formally state the problem and briefly discuss our main contributions.

\subsection{Network Model}
In this paper, we consider a network modeled by a random geometric graph $G(\lambda, R)$, whose nodes are deployed in a region $\mathcal{S}$ according to a Poisson point process with density $\lambda$. There is an edge between each pair of nodes if their Euclidean distance is less than $R$.
We assume that $\mathcal{S}$ is a circular region with diameter $D$ and centered at the origin $\textbf{0}$. However, the results presented in this paper can be extended to other types of deployment regions with minimal changes.
Initially, all the nodes carry the same amount of load $l$, and have the same capacity $c=\alpha l$, where $\alpha \geq 1$ is the \emph{tolerance parameter}. While the load of each node may change over time, the capacity remains the same. A node is called ``healthy" if it carries a load less than or equal to its capacity.

\emph{Connected vs. Disconnected Graphs:} By definition, in a connected network, there exists a path between any two arbitrary nodes in the network. For $G(\lambda, R)$, connectivity is only guaranteed when $\lambda\rightarrow \infty$. In practice, however, the probability of connectivity could be arbitrarily close to 1 if $\lambda$ is chosen large enough. Note that $G(\lambda, R)$ defines a probability space with a sample space consisting of all possible realizations of $G(\lambda, R)$.
For finite values of $\lambda$, let $G_c(\lambda, R)$ be the \emph{connected} subspace of the larger probability space $G(\lambda, R)$, formed by all the connected realizations of $G(\lambda, R)$.
In our analysis, whenever connectivity is needed, we will consider $G_c(\lambda, R)$. In simulations, however, it is extremely time-consuming to check the connectivity of each realization. Therefore, in order to have a connectivity probability close to 1,
we assume $\lambda$ is chosen such that $\lambda \pi R^2\geq 6$. Hence, the probability of a node being isolated, which accounts for the dominant term in the probability of disconnectivity, is upper-bounded as $\exp(-\lambda \pi R^2)\leq \exp(-6)=2.5\times 10^{-3}$ \cite{Bettstetter02}.

\subsection{Initial Disturbance and Propagation of Failures}
\begin{figure}[t]
\centering
{\includegraphics[width =4 in , height=3.9 in]{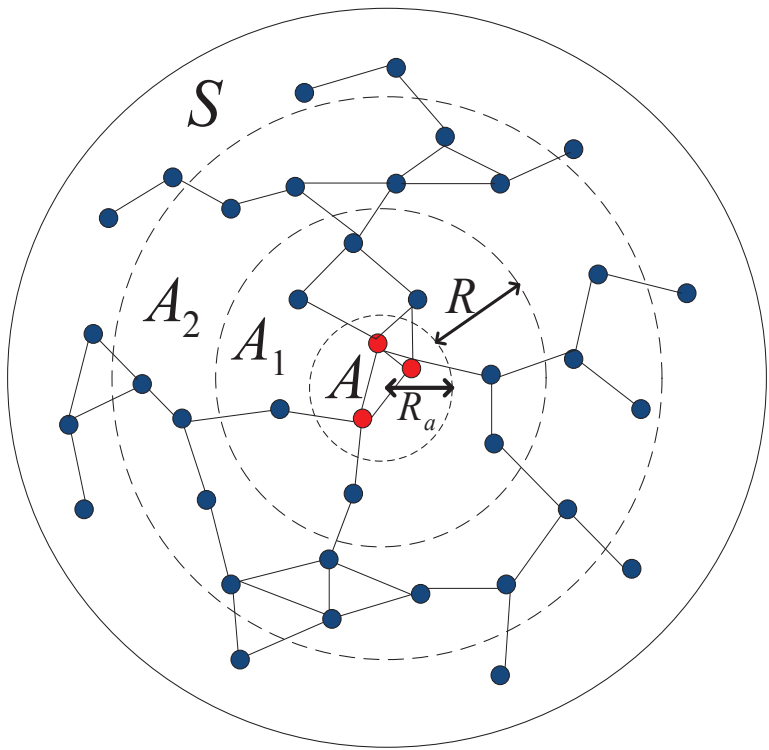}}
\caption{Dish attack (shaded area) in a random geometric graph.}
\label{fig:RGG}
\end{figure}

A \emph{dish attack} on $\mathcal{S}$ is modeled by a circle $\mathcal{A}$ of radius $R_a< D/2$ centered at the origin. This is shown in Fig. \ref{fig:RGG}.
After the attack, all the nodes located at a distance $r<R_a$ from the center of attack will fail, and their load will be redistributed to their neighbors, which in turn may lead to a propagation of failures throughout the network. We assume that a dish attack only affects the nodes inside the dish, not the ones located on its border at $r=R_a$.
We focus on the set of conditions under which a cascading failure is realized, and study the corresponding damage caused by such a cascade.
We assume the following model for the propagation of failures. At any stage of cascade, when a node fails, its load will be redistributed \emph{equally} among its \emph{healthy} neighbors. A node that carries a total load greater than its capacity will fail.

\subsection{Quantifying the Damage}
The number of failures at each stage of load redistribution is clearly a random variable (r.v.). In order to quantify the impact of an attack on the network, we use the total number of failures outside the attacked region, caused by a limited dish attack. Let $F$ denote this number. We define \emph{failure ratio} as
\begin{align}
f\triangleq\frac{F}{|\mathcal{S}\setminus \mathcal{A}|},
\end{align}
where $|\mathcal{S}\setminus \mathcal{A}|$ is the total number of nodes outside the attack region, including the nodes on $\mathcal{A}$'s boundary.
We use the average value of the random variable $f$ taken over all realizations of $G(\lambda, R)$, denoted by $\bar{f}$, to measure the impact of an attack. We are particularly interested in the variation of $\bar{f}$ with the tolerance parameter $\alpha$.
Fig. \ref{fig:samplealphas} shows $\bar{f}$ versus $\alpha$ for a typical dish attack on a network where $R_a=R=0.1$ and $D=1$, for different values of $\lambda$.

\subsection{Main Results}
As noted above, we discussed the insufficiency of the existing asymptotic analyses when applied to finite-size applications.
In this paper, we study the reaction of a finite-size network $G(\lambda, R)$ to a dish attack of an arbitrary radius $R_a$ by providing analytical results for $\bar{f}$ in terms of $\lambda$, $R$, $R_a$, and the most important parameter, $\alpha$.
Finding the exact value of $\bar{f}$ in the finite regime could be very difficult and, if found, it may very well result in computationally intensive, if not intractable, arguments. Instead, we focus on deriving bounds with manageable computational complexity that help us understand the variations of $\bar{f}$ as the network parameters change. We summarized our main contributions as follows.

\begin{figure}[t]
\centering
{\includegraphics[width =4 in , height=3.4 in]{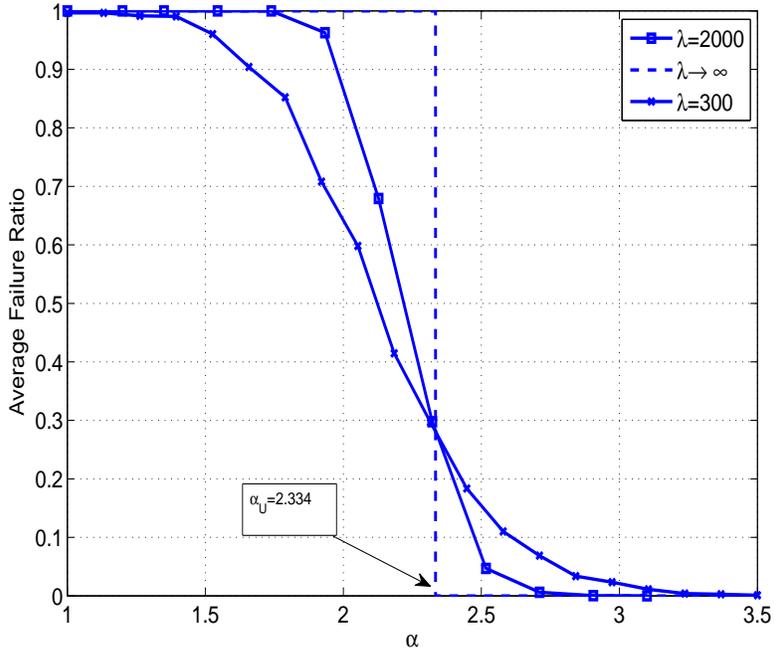}}
\caption{Average failure ratio versus $\alpha$ in both finite and large-scale networks when $R_a=R=0.1$ and $D=1$. As $\lambda$ grows larger, a threshold behavior with respect to the tolerance parameter is observed. }
\label{fig:samplealphas}
\end{figure}

\begin{itemize}
\item We start by investigating the first few stages of load redistribution after a dish attack, particularly finding the load redistributed to nodes in $\mathcal{A}_1$ immediately after the attack. We extend this analysis to obtain an upper bound on the average failure ratio $\bar{f}$, which especially helps us choose an appropriate value of $\alpha$ to avoid a cascade.

\item In order to derive a lower bound, we consider a favorable scenario for absorbing the load redistributed from failed nodes, by assuming a desirable network topology and full node cooperation. It will be shown that, even in such an optimistic scenario, the chance to stop a cascade becomes smaller and smaller as the failures propagate through the network. This leads to a lower bound on $\bar{f}$.

\item The two bounds together provide us with insights into the speed and extent of a failure cascade through the network. Our analysis reveals the critical role of the first few stages of load redistribution in preventing a cascade. In other words, our results indicate that if a spread of failures is not contained immediately or within the first few stages, a cascade of failures would most likely bring down a large portion of the network.

\item As seen from Fig. \ref{fig:samplealphas}, the failure ratio changes rather quickly over a short interval of $\alpha$. It will be shown that this interval diminishes to zero as $\lambda$ increases, indicating a phase transition phenomenon. Investigating the proposed upper bound on $\bar{f}$ as $\lambda \rightarrow \infty$ reveals the existence of a threshold value of $\alpha$, denoted as $\alpha_U$, such that $\bar{f}=1$ if $\alpha< \alpha_U$, and $\bar{f}=0$ if $\alpha \geq \alpha_U$. We will derive $\alpha_U$ in terms of other network parameters.
\end{itemize}

\section{Preliminaries}\label{sec:prelim}
Here, we provide some notation and preliminaries required for the analysis. We denote the number of nodes in the attack region $\mathcal{A}$ by $a$. Note that $a$ is a Poisson r.v. with parameter
\begin{align}
\bar{a}=\delta_a^2=\lambda\pi R_a^2 . 
\end{align}
It makes sense to assume that a dish attack is large enough to affect at least one node, for which we assume $\bar{a}=\lambda\pi R_a^2  \geq 3$ in this paper, which yields Pr$(a \geq 1)> 0.95$.

Consider the rings (annuli) of width $R$ around the attacked region, as depicted in Fig. \ref{fig:RGG}. For $i\geq 1$, we denote an annulus with inner radius $R_{i-1}=R_a+(i-1)R$ and outer radius $R_i=R_a+iR$ by $\mathcal{A}_i$, and the set of nodes in $\mathcal{A}_i$ by $A_i$. We denote the cardinality of $A_i$ by $a_i$. Note that $a_i$, which is the number of nodes in the ring $\mathcal{A}_i$, is simply a Poisson random variable with parameter
\begin{align}\label{eq:ai}
\bar{a}_i=\delta_{a_i}^2=\lambda \pi(R_i^2-R_{i-1}^2).
\end{align}
The following lemmas will help us in our sequential analysis, whose proofs are provided in the appendix.

\begin{lem}\label{lem:gaus}
Let $\bar{a}=\lambda\pi R_a^2  \geq 3$ and $\lambda \pi R^2\geq 6$, as assumed in this paper. We then have
\begin{align}
\bar{a}_i > 14, \quad i\geq 1.
\end{align}
\end{lem}
Since $\bar{a}_i$ is greater than 10, Lemma \ref{lem:gaus} implies that the Poisson r.v. $a_i$ could be well approximated by a Gaussian r.v. for $i\geq 1$ \cite{RossProbability}. We will make it clear when we use this assumption in the later analysis.
Consider two circles, one with radius  $r_1$ centered at a distance $a$ from the origin, and the other one with radius $r_2$ centered at a distance $b$.
We denote by $\mathcal{I}(a,r_1,b,r_2)$ the intersection region of these two circles, while we use $I(a,r_1,b,r_2)$ to represent the area of this region, which could be obtained as \cite{RossProbability}
\begin{align}
\notag I(a,r_1,b,r_2))=& r_2^2 \cos^{-1}\big(\frac{(b-a)^2+r_2^2-r_1^2}{2|b-a|r_2}\big)+r_1^2 \cos^{-1}\big(\frac{(b-a)^2+r_1^2-r_2^2}{2|b-a|r_1}\big)\\
&-1/2 \sqrt{(-|b-a|+r_1+r_2)(|b-a|+r_2-r_1)(|b-a|-r_2+r_1)(|b-a|+r_2+r_1)}.
\end{align}

\begin{figure}[t]
\centering
{\includegraphics[width =3.4 in , height=3.4 in]{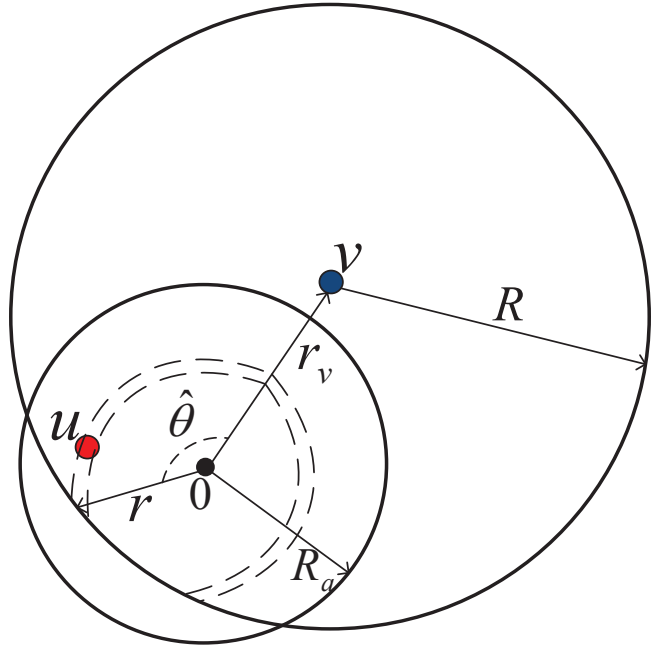}}
\caption{Figure shows the intersection area $\mathcal{I}(0,R_a, r_v, R)$. We are interested in finding $\psi(r)$, the PDF of the distance $r$ of a node $u$ located randomly and uniformly on $\mathcal{I}(0,R_a, r_v, R)$.}
\label{fig:locationdist}
\end{figure}

\begin{lem}\label{lem:dist}
Let $u$ be a node located randomly and uniformly on $\mathcal{I}(0,R_a, r_v, R)$ with $r_v\geq R_a$, as shown in Fig. \ref{fig:locationdist}. Also let $r$ be the random variable representing $u$'s distance from the center of attack (i.e., the origin). Then the probability distribution function (PDF) of $r$ is given as
\begin{equation}\label{eq:phi}
\psi(r)= \left\{\begin{aligned}
        &\frac{2r}{I(0,R_a, r_v, R)} \arccos{\big(\frac{r_v^2-R^2+r^2}{2r_vr}\big)}& \quad \textrm{if}\quad r+r_v> R\\
&\frac{2\pi r}{I(0,R_a, r_v, R)}& \quad \textrm{if}\quad r+r_v\leq R.
       \end{aligned}
 \right.
\end{equation}
\end{lem}

\section{Upper Bound on the Failure Ratio}\label{sec:upper}
In this section, we turn our attention to a \emph{necessary condition} for having a cascade. A cascade of failures is possible only if at least one node outside the attack region fails due to the load redistribution. Otherwise, if the load of the attacked nodes in $\mathcal{A}$ is completely absorbed by the rest of the network, the propagation of failure does not occur. By finding the probability of this event, we could derive an upper bound on the average failure ratio.

We start our analysis for the finite-size networks by investigating the load received by nodes outside the attack region, immediately after the attack. This is the load received by immediate neighbors of the attacked nodes after the very first load redistribution.
Note that this load is a random variable. Also recall that the neighbors of the attacked region are all located in $\mathcal{A}_1$. We will first find the mean and standard deviation of the load received by these nodes.
Having the statistics of this random variable, we then show that its distribution could be well approximated by a Gaussian random variable.
Using such an approximation, we then find the probability of an overload for the nodes in $\mathcal{A}_1$, which later helps us find an upper bound on the average failure ratio. Recall that ``average" here stands for an average taken over all graph realizations. Before presenting the main result of this section, we need to state the following lemmas, whose proofs could be found in the appendix.

\begin{lem}\label{lem:poi}
Let $d_u$ be a Poisson random variable with density $\lambda_u$. We then have
\begin{align}
&E[\frac{1}{d_u} \mid d_u>0]= \frac{e^{-\lambda_u }g(\lambda_u)}{1-e^{-\lambda_u}},\\
&E[\frac{1}{d_u^2} \mid d_u>0]= \frac{e^{-\lambda_u }}{1-e^{-\lambda_u}}\int_{-\infty}^{\lambda_u} \frac{1}{x} g(x) dx, \label{eq:poi2}
\end{align}
where
\begin{align}\label{eq:g}
g(x)=\sum_{k=1}^{\infty} \frac{1}{k} \frac{x^k}{k!}=\int_{-\infty}^{x} \frac{e^{z}-1}{z} dz.
\end{align} 
\end{lem}

\begin{figure}[t]
\centering
{\includegraphics[width =3.5 in , height=2.8 in]{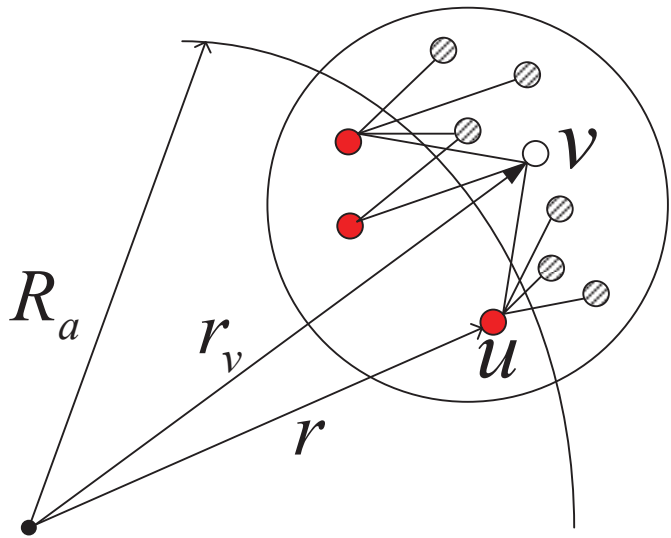}}
\caption{Figure shows the setting in the proof of Lemma \ref{lem:i}. Black nodes are neighbors of $v$ in the attacked region.}
\label{fig:avedeganal}
\end{figure}

\begin{lem}\label{lem:i}
Consider a node $v$ located at a distance $r_v\in[R_a, R_a+R)$ from the center of attack. Also consider a node $u$, a neighbor of $v$, located inside the attacked region at a distance $r<R_a$ from the center, as shown in Fig. \ref{fig:avedeganal}.
The average load $l_u$ redistributed to $v$ from $u$ can be obtained as
\begin{align}\label{eq:lgivenr}
E[l_u|r]=\frac{e^{-\lambda J(r) }g(\lambda J(r))}{1-e^{-\lambda J(r)}} \triangleq h^{(1)}(r),
\end{align}
where $g(\cdot)$ is defined in (\ref{eq:g}) and $J(r)=\pi R^2- I(r,R,0,R_a)$. Moreover, we have
\begin{align}\label{eq:l2givenr}
E[l_u^2|r]=\frac{e^{-\lambda J(r) }}{1-e^{-\lambda J(r)}} \int_{-\infty}^{\lambda J(r)} \frac{g(x)}{x}  dx\triangleq h^{(2)}(r).
\end{align}
If $u$ is located randomly and uniformly on $\mathcal{I}(r_v,R,0,R_a)$, we have
\begin{align}
E[l_u]= \int_{r_v-R}^{R_a} h^{(1)}(r)\times \psi(r) \ dr,\label{eq:aveli}\\
E[l_u^2]= \int_{r_v-R}^{R_a} h^{(2)}(r) \times \psi(r)\ dr.\label{eq:aveli2}
\end{align}
If $R-r_v\geq R_a$, we obtain $J(r)=J^*\triangleq\pi R^2- \pi R_a^2$, and (\ref{eq:aveli}) and (\ref{eq:aveli2}) could be reduced to
\begin{align}\label{eq:lihalf}
\notag &E[l_u]=\frac{e^{-\lambda J^* }g(\lambda J^*)}{1-e^{-\lambda J^*}},\\
&E[l_u^2]=\frac{e^{-\lambda J^* }}{1-e^{-\lambda J^*}} \int_{-\infty}^{\lambda J^*} \frac{g(x)}{x} dx.
\end{align}
Finally, given $E[l_u]$ and $E[l_u^2]$, the variance $\sigma_{l_u}^2$ is given as
\begin{align}\label{eq:sigmali}
\sigma_{l_u}^2=E[l_u^2]-E^2[l_u].
\end{align} 
\end{lem}

The following theorem applies the results of Lemmas \ref{lem:dist}, \ref{lem:poi}, and \ref{lem:i} to find the mean and variance of the load redistributed to a node at distance $r_v$ from the center of attack, right after the attack.
\begin{thm}\label{thm:aveload}
Consider a node $v$ located at a distance $r_v\in[R_a, R_a+R)$ from the center of attack. Let $L_v$ be the load redistributed to $v$ by its neighbors inside the attacked region. We then have
\begin{align}
&E[L_v]=\lambda I(r_v,R, 0,R_a) \int_{r_v-R}^{R_a} h^{(1)}(r) \times \psi(r) \ dr,\label{eq:aveload}\\
&\sigma_{L_v}^2=\lambda I(r_v,R, 0,R_a)\times \sigma_{l_u}^2,\label{eq:sigmaload}
\end{align}
where $h^{(1)}(r)$ is defined in (\ref{eq:lgivenr}), and $\sigma_{l_u}^2$ is given by (\ref{eq:sigmali}). If $R-r_v\geq R_a$, (\ref{eq:aveload}) and (\ref{eq:sigmaload}) are reduced to
\begin{align}\label{eq:Lhalf}
\notag &E[L_v]=\lambda\pi R_a^2\times \frac{e^{-\lambda J^* }g(\lambda J^*)}{1-e^{-\lambda J^*}},\\
&\sigma_{L_v}^2=\lambda\pi R_a^2\times \sigma_{l_u}^2,
\end{align}
where $g(\cdot)$ is given by (\ref{eq:g}), and $J^*=\pi R^2- \pi R_a^2$. 
\end{thm}

Now that we have the mean and variance of $L_v$, an approximation of $L_v$'s PDF could be obtained using the central limit theorem as follows. Note that
\begin{align}
L_v=\sum_{u=1}^N l_u,
\end{align}
where $N$ is the number of nodes inside $\mathcal{I}(r_v,R,0,R_a)$, a Poisson r.v. with mean $\lambda I(r_v,R,0,R_a)$. Given that $N=n$, the nodes $u=1,...,n$ would be distributed randomly and independently on $\mathcal{I}(r_v,R,0,R_a)$, making $l_u$'s i.i.d. random variables. Therefore, for large values of $n$, the central limit theorem asserts that the probability distribution of $L_v$ is well-approximated by a Gaussian random variable. In practice, however, $n\geq 5$ is large enough to ensure a PDF very close to the normal random variable \cite{RossProbability}. The following corollary is a formal statement of what we just explained.

\begin{cor}\label{cor normala:pprox}
If $\lambda\times I(r_v,R, 0,R_a)>>1$, the load received by a node $v$ at $r_v$ could be approximated by a Gaussian r.v. $L_v\sim \mathcal{N}(\bar{L}_v, \sigma_{L_v}^2)$, where $\bar{L}_v$ and $\sigma_{L_v}^2$ are given by (\ref{eq:aveload}) and (\ref{eq:sigmaload}), respectively. In particular, we have
\begin{align}\label{eq:normalapp}
\textrm{Pr}\{v\ \textrm{fails}\}=\textrm{Pr}\{L_v> \alpha-1\}\approx 1-\Phi(\frac{\alpha-1-\bar{L}_v}{\sigma_{L_v}}),
\end{align}
where $\Phi(\cdot)$ is the cumulative distribution function (CDF) of the standard normal distribution with mean 0 and variance 1. 
\end{cor}

When $\lambda I(r_v,R,0,R_a)$ is small due to either $\lambda$ or $I(r_v,R,0,R_a)$, the load received by $v$ becomes very small. Since the Gaussian-approximated value for $L_v$ also becomes small in this case, the error in the approximation becomes negligible. The following theorem employs this fact along with Corollary \ref{cor normala:pprox} to find the probability of survival for the nodes in $\mathcal{A}_1$ after the very first round of load redistribution.

\begin{thm}\label{th:loadprob}
Let $v$ be a node located randomly and uniformly on $\mathcal{A}_1$. Also let $p_1$ be the probability that the load received by $v$ is less than or equal to $\alpha-1$, i.e., $p_1\triangleq\textrm{Pr}\{L_{v}\leq \alpha-1\}$. Then, $p_1$ is obtained as
\begin{align}\label{eq:loadprob}
p_1\approx \int_{R_a}^{R_a+R} \Phi\big(\frac{\alpha-1-\bar{L}_v}{\sigma_{L_v}}\big)\times \frac{2\pi r_v}{|\mathcal{A}_1|}\ dr_v.
\end{align}
Note that $\bar{L}_v$ and $\sigma_{L_v}$ are functions of $r_v$, given by Theorem \ref{thm:aveload}. 
\end{thm}

Using the finding of Theorem \ref{th:loadprob}, an upper bound  on the average failure ratio can be obtained for finite values of $\lambda$.
\begin{thm}\label{th:upper}
The average failure ratio due to a dish attack of radius $R_a$ is upper-bounded as
\begin{align}\label{eq:f}
\bar{f}\leq 1-e^{-\lambda_1(1-p_1)},
\end{align}
where
\begin{align}
\lambda_1=\lambda \pi\big((R_a+R)^2-R_a^2\big)
\end{align}
is the density of nodes in $\mathcal{A}_1$, and $p_1$ is given by Theorem \ref{th:loadprob}. 
\end{thm}

\begin{figure}[t]
\centering
{\includegraphics[width =3.5 in , height=2.8 in]{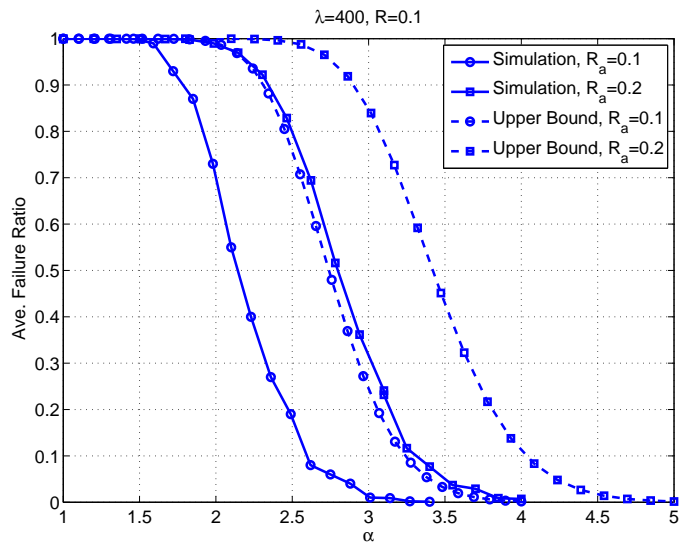}}
\caption{The upper bound from Theorem \ref{th:upper} against the simulation results,  provided for a dish attack on a network $G(\lambda=400, R=0.1)$.}
\label{fig:upper}
\end{figure}

Fig. \ref{fig:upper} depicts the upper bound from Theorem \ref{th:upper} for different values of network parameters, where we also include the simulation results for the exact value of $\bar{f}$. As seen, the proposed upper bound is especially helpful when it comes to picking a value of $\alpha$ to avoid a cascade. For example, for the network $G(\lambda=400, R=0.1)$, the upper bound suggests that $\alpha=3$ is a good choice to contain dish attacks of radius $R_a=0.1$ or smaller.

\section{Asymptotic Analysis of Upper Bound and Threshold Behavior of Failure Ratio}\label{sec:asympupper}
While Theorem \ref{th:upper} provides an upper bound on the failure ratio in finite-size networks, an asymptotic analysis of the upper bound could provide intuition regarding the behavior of a large network under attacks. As we will see, such an analysis reveals the threshold behavior of the failure ratio in terms of the tolerance parameter. For the case with $\lambda \rightarrow \infty$, it could be shown that as the tolerance parameter increases above 1, the failure ratio drops from 1 to 0 at a critical value of the tolerance parameter. We will find such critical value, which could be very helpful when studying large networks' robustness to cascades. We start our analysis by finding what happens to the load $L_v$ in Theorem \ref{thm:aveload} when $\lambda\rightarrow \infty$. In this section, in order to explicitly show the dependence of $L_v$ on $r_v$, we use the notation $L(r_v)$ instead of $L_v$ for the load received by node $v$ located at $r_v$. This slight modification will prove helpful in understanding the analysis.

\begin{thm}\label{th:Lasymp}
Consider a dish attack of radius $R_a$ applied to a network $G(\lambda, R)$. Let $L(r_v)$ be the load received by a node $v$ located at a distance $r_v\in[R_a, R_a+R)$ from the center of attack, right after the attack. When $\lambda\rightarrow \infty$, $L(r_v)$ is no longer a random variable, and given as
\begin{align}\label{eq:Lasymp}
\underset  {\lambda\rightarrow \infty} {L(r_v)} \rightarrow 2 \int_{r_v-R}^{R_a} \frac{ r } {J(r)} \arccos{\big(\frac{r_v^2-R^2+r^2}{2r_vr}\big)} dr,
\end{align}
where $J(r)=\pi R^2- I(r,R, 0,R_a)$. 
\end{thm}

Having the asymptotic value of $L(r_v)$ from Theorem \ref{th:Lasymp}, a sufficient condition for a cascade of failures in the asymptotic case could be obtained as below.

\begin{thm}\label{th:cascade}
Consider a healthy node $v$ located at $r_v\geq R_a$ after a dish attack of radius $R_a$ on $G(\lambda, R)$ where $\lambda \rightarrow \infty$.
Let
\begin{align}\label{eq:alphaU}
\alpha_U\triangleq 1+ L(R_a)=1+ 2 \int_{R_a-R}^{R_a} \frac{ r } {J(r)} \arccos{\big(\frac{R_a^2-R^2+r^2}{2R_ar}\big)} dr.
\end{align}
If $\alpha<\alpha_U$ and all the nodes located at $r<r_v$ have failed, then $v$ will fail as well. Hence, a cascade of failures occurs throughout the network, resulting in $\bar{f}=1$. 
\end{thm}

The following theorem combines the sufficient condition from Theorem \ref{th:cascade} with a necessary condition for a cascade, proving a threshold behavior for the average failure ratio in the asymptotic regime.

\begin{thm}\label{th:upperasymp}
Consider a dish attack of radius $R_a$ applied to a network $G(\lambda, R)$ where $\lambda \rightarrow \infty$.
Let $\alpha_U$ be the value of $\alpha$ given by (\ref{eq:alphaU}).
Then, $\bar{f}=0$ if $\alpha\geq\alpha_U$, and $\bar{f}=1$ if $\alpha<\alpha_U$. 
\end{thm}

\begin{figure}[t]
\centering
{\includegraphics[width =3.5 in , height=2.8 in]{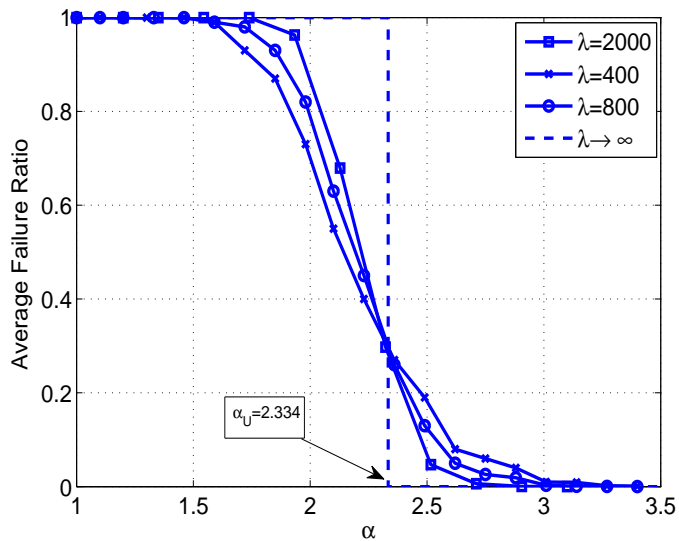}}
\caption{Figure illustrates the threshold behavior of failure ratio in terms of $\alpha$. The dashed curve shows the threshold value $\alpha_U$ given by Theorem \ref{th:cascade}. Simulation results (solid lines) are presented for a dish attack of radius $R_a=0.1$ on $G(\lambda, R=0.1)$ for different values of $\lambda$.}
\label{fig:asymp}
\end{figure}

Fig. \ref{fig:asymp} demonstrates the evolution of the average failure ratio $\bar{f}$ as $\lambda$ grows larger. It also shows the value of $\alpha_U$ given by Theorem \ref{th:cascade} for the asymptotic case. As it can be seen, a phase transition around $\alpha_U$ becomes clear as $\lambda$ increases.

\section{Lower Bound on the Failure Ratio}\label{sec:lower}
In this section, we derive a lower bound on the failure ratio by analyzing a \emph{sufficient condition} for the propagation of failures throughout the network. This condition is based on the fact that if a cascade cannot be stopped in the presence of a) full cooperation between nodes, and b) the most favorable connectivity condition, then for sure it cannot be stopped without them. We first provide the lower bound for finite-size networks.

\begin{thm}\label{th:lowerbound}
Consider the connected subspace $G_c(\lambda, R)$ of the probability space $G(\lambda, R)$, introduced in Section \ref{sec:model}. Suppose that a dish attack of radius $R_a$ is applied to $G_c(\lambda, R)$. Also let $q$ denote the ratio $R_a/R$. If
\begin{align}
\alpha< 3/2+q,
\end{align}
then we have
\begin{align}\label{eq:lowerbound}
\bar{f}\geq e^{-\bar{a}} \sum_{k=1}^{\infty} \Phi\big(\frac{\frac{k}{\alpha-1}-\bar{a}_1}{\sqrt{\bar{a}_1}}\big) \frac{\bar{a}^k}{k!},
\end{align}
where $\bar{a}=\lambda\pi R_a^2$, $\bar{a}_1=\lambda \pi[(R_a+R)^2-R_a^2]$, and $\Phi(\cdot)$ is the CDF of a standard normal distribution. 
\end{thm}

In practice, the summation in (\ref{eq:lowerbound}) needs to be calculated only for $\lfloor2\bar{a}\rfloor$ or $\lfloor3\bar{a}\rfloor$ terms. For that, let us consider the terms after $k=\lfloor3\bar{a}\rfloor$ in (\ref{eq:lowerbound}).
We have
\begin{align}\label{eq:3abound}
\notag e^{-\bar{a}} \sum_{\lfloor3\bar{a}\rfloor+1}^{\infty} \Phi\big(\frac{\frac{k}{\alpha-1}-\bar{a}_1}{\sqrt{\bar{a}_1}}\big) \frac{\bar{a}^k}{k!}&\overset{(a)}{\leq} e^{-\bar{a}} \sum_{\lfloor3\bar{a}\rfloor+1}^{\infty} \frac{\bar{a}^k}{k!}= Pr\{a\geq \lfloor3\bar{a}\rfloor+1\} \leq Pr\{a\geq 3\bar{a}\}\\
\notag &\overset {(b)}{\leq} \frac{e^{-\bar{a}} (e\times \bar{a})^{3\bar{a}}}{(3\bar{a})^{3\bar{a}}}=(\frac{e^2}{3^3})^{\bar{a}}=(0.2737)^{\bar{a}}\overset{(c)}{\leq} (0.2737)^3\\
&=0.0205,
\end{align}
where (a) holds because we have $\Phi(\cdot)\leq 1$, (b) is obtained by applying the Chernoff bound to the Poisson tail probability \cite{Mitzenmacher052} \footnote{An upper Bound for the tail probability of a Poisson random variable  $X \sim \text{Poi}(\lambda)$ can be derived using a Chernoff bound argument \cite{Mitzenmacher052}: $Pr\{X\geq x\}\leq \frac{e^{-\lambda}(e\lambda)^x}{x^x}$, for $x>\lambda$.}, and (c) is due to our assumption of $\bar{a}\geq 3$ in this paper. The small value 0.0205, when compared to 1, could be safely omitted for practical purposes.

Before proving Theorem \ref{th:lowerbound}, we need to state a few lemmas. Recall the rings $\mathcal{A}_i$, $i\geq1$, in Fig. \ref{fig:RGG}. A cascade of failures, at each stage of its progress, goes through one of these rings. Let us look at how the failure propagates after the attack. After an attack on $\mathcal{A}$, all the nodes that may potentially fail in the next step are the neighbors of $\mathcal{A}$ located in $\mathcal{A}_1$. If some of the nodes in $\mathcal{A}_1$ fail, the next step of propagation includes some nodes in $\mathcal{A}_1$ and $\mathcal{A}_2$. In general, if the failures have already been spread trough $\mathcal{A}_1,..., \mathcal{A}_i$, potential failures of the next step are all in $(\mathcal{A}_1\cup ...\cup \mathcal{A}_i) \cup \mathcal{A}_{i+1}$. We know that $a_i$'s are Poisson random variables with parameter $\bar{a}_i$'s given by (\ref{eq:ai}). Lemmas \ref{lem:firststep} and \ref{lem:ratio} below establish a connection between $\bar{a}_i$'s and $\alpha$. We use this connection later in Lemma \ref{lem:Pr} to prove a useful property in finding the lower bound. The proofs of all lemmas can be found in the appendix.

\begin{lem}\label{lem:firststep}
Let $q=R_a/R$. Given that $a=a_0$, if $\alpha-1< 1/2+q$, we have
\begin{align}
\frac{\bar{a}_2}{\bar{a}_1}< \frac{\alpha}{\alpha-1}.
\end{align} 
\end{lem}

\begin{lem}\label{lem:ratio}
For $i\geq 2$, we have
\begin{align}\label{eq:ratiofirst}
\frac{\bar{a}_{i+1}}{\bar{a}_i} \leq \frac{\bar{a}_i}{\bar{a}_{i-1}}.\
\end{align}
Particularly, if $\alpha-1< 1/2+q$ as in Lemma \ref{lem:firststep}, we have
\begin{align}\label{eq:ratiosec}
\frac{\bar{a}_{i+1}}{\bar{a}_i} \leq \frac{\bar{a}_2}{\bar{a}_1}< \frac{\alpha}{\alpha-1},
\end{align}
for $i\geq 2$. 
\end{lem}

\begin{remark}
\textbf{Gaussian approximation for $a_i, i\geq1$:} Recall Lemma \ref{lem:gaus} and the discussion afterwards where we explained how our assumptions for $\lambda \pi R_a^2$ and $\lambda \pi R^2$ lead to $\bar{a}_i>14$, for $i\geq1$. This means that the Poisson r.v. $a_i$ is well approximated by a Gaussian r.v. with the same mean and variance as $a_i$ given by (\ref{eq:ai}). We use this approximation in proving the following lemma.
\end{remark}

\begin{lem}\label{lem:Pr}
Given that $a=a_0$ and $\alpha-1<1/2+q$, the following property holds for $a_i, i\geq 0$.
\begin{align}
\textrm{Pr}\{a_0+a_1+...+a_i > a_{i+1}(\alpha-1)\} \geq \textrm{Pr}\{a_0> a_1(\alpha-1)\}.
\end{align}
\end{lem} 

Using the preliminary results stated above, we obtain the following theorem regarding a total-failure cascade in a finite-size network, whose proof is given in the appendix.

\begin{lem}\label{lem:a0fails}
Suppose a dish attack of radius $R_a$ is applied to $G_c(\lambda, R)$. Given that $a=a_0$, if $\alpha-1< 1/2+q$, the probability that all the nodes fail is lower-bounded by $\textrm{Pr}\{a_0> a_1(\alpha-1)\}$.
\end{lem}

Now we are ready to prove Theorem \ref{th:lowerbound}.

\emph{Proof of Theorem \ref{th:lowerbound}:} Let us define an identity random variable $X_{a}$ as
\[X_{a}=\left\{
\begin{array}{l l}
1 & \quad \textrm{if $a$ causes a total failure,}\\
0 & \quad \textrm{otherwise}.
\end{array} \right.\]
We can write
\begin{align}
\bar{f}\geq \sum_{a\geq0} X_{a} \textrm{Pr}\{X_{a}=1\} & \overset{(a)}{=} \sum_{k\geq 1} \textrm{Pr}\{a>(\alpha-1)a_1 | a=k\}\times \textrm{Pr}\{a=k\}\\
&\overset{(b)}{=}e^{-\bar{a}} \sum_{k=1}^{\infty} \Phi\big(\frac{\frac{k}{\alpha-1}-\bar{a}_1}{\sqrt{\bar{a}_1}}\big) \frac{\bar{a}^k}{k!},
\end{align}
where $(a)$ is due to Lemma \ref{lem:a0fails}, and $(b)$ follows from the Gaussian distribution of $a_1$ and Poisson distribution of $a$. \qedb

\begin{figure}[t]
\begin{center}
\subfigure[Dish attack of radius $R_a=0.1$ applied to $G(\lambda=400, R=0.1)$.]
{\includegraphics[width =4.5 in , height=3.5 in]{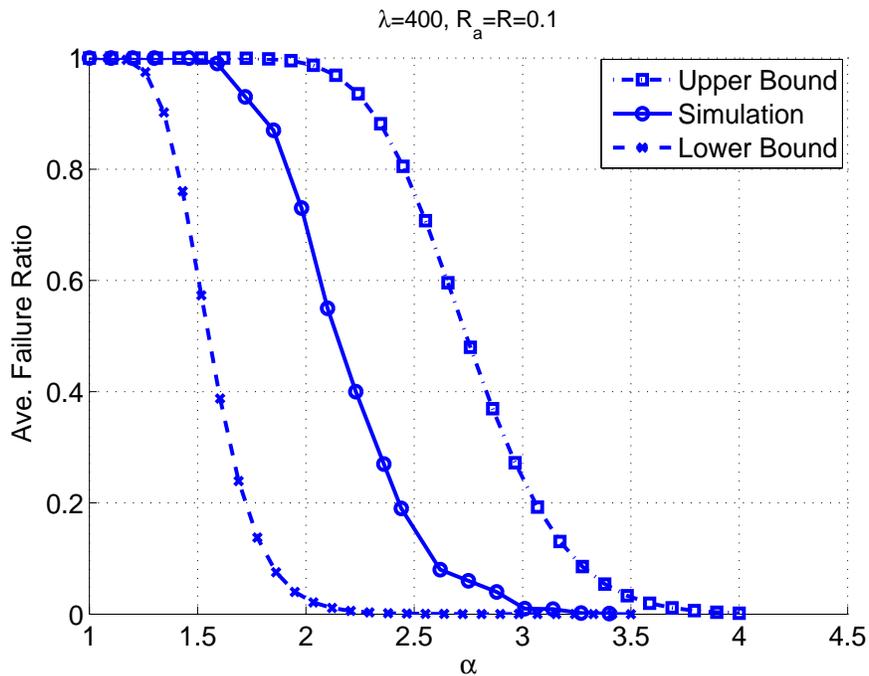} \label{fig:lowa}}
\subfigure[Dish attack of radius $R_a=0.2$ applied to $G(\lambda=400, R=0.1)$.]
{\includegraphics[width =4.5 in , height=3.5 in ]{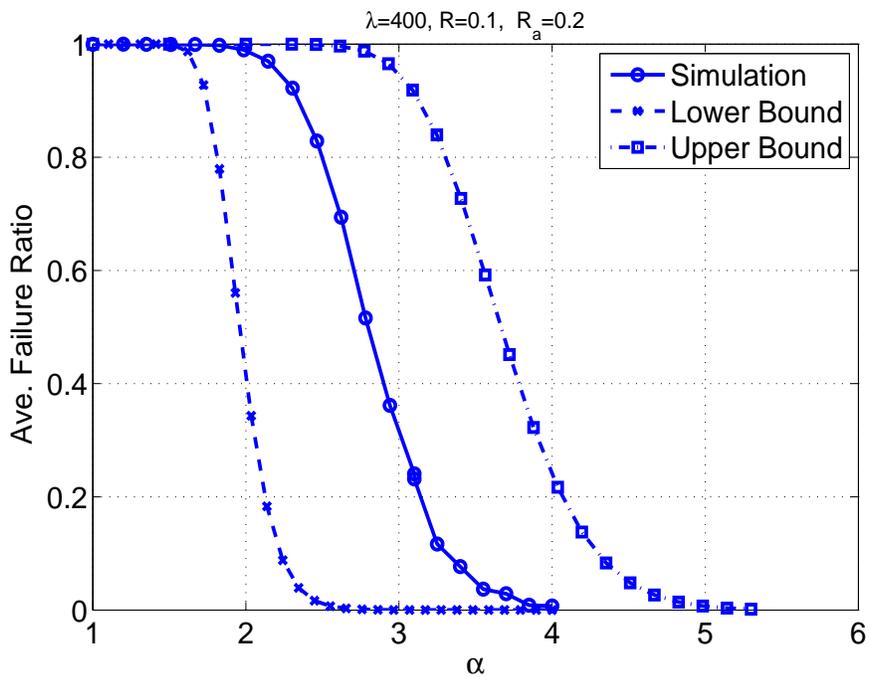} \label{fig:lowb}}
\end{center}
\caption{Simulation results for the average failure ratio versus $\alpha$, along with the lower bound of Theorem \ref{th:lowerbound} and upper bound of Theorem \ref{th:upper}. The results are shown for dish attacks of radii $R_a=0.1$ and $R_a=0.2$, respectively, applied to $G(\lambda=400, R=0.1)$. Network diameter $D$ is set to 1.}\label{fig:low}
\end{figure}

Fig. \ref{fig:low} depicts the lower bound from Theorem \ref{th:lowerbound} along with the simulation result for the average failure ratio. The upper bound from Theorem \ref{th:upper} is also shown for comparison. As we see, the two bounds together successfully predict the interval within which the failure ratio decreases from 1 to 0.

\subsection{Asymptotic Analysis of the Lower Bound}
Here, we look at the lower bound obtained in the previous section from an asymptotic  point of view. As $\lambda$ grows very large, similar to what was observed for the upper bound in Section \ref{sec:asympupper}, the lower bound takes the shape of a step function. That is, there exists a value of $\alpha$, denoted as $\alpha_L$, such that $\bar{f}$ takes the value of 1 for $\alpha< \alpha_L$, and it takes the value of 0 for $\alpha\geq \alpha_L$. The following theorem derives the value of $\alpha_L$.

\begin{thm}\label{th:alphaL}
Consider the probability space  $G(\lambda, R)$ when $\lambda\rightarrow \infty$. Suppose that a dish attack of radius $R_a$ is applied to $G(\lambda, R)$.
Also let $q$ denote the ratio $R_a/R$. If
\begin{align}
\alpha< \alpha_L \triangleq 1+ \frac{q^2}{1+2q},
\end{align}
all the nodes would fail. 
\end{thm}

\begin{figure}[t]
\centering
{\includegraphics[width =4 in , height=3.4 in]{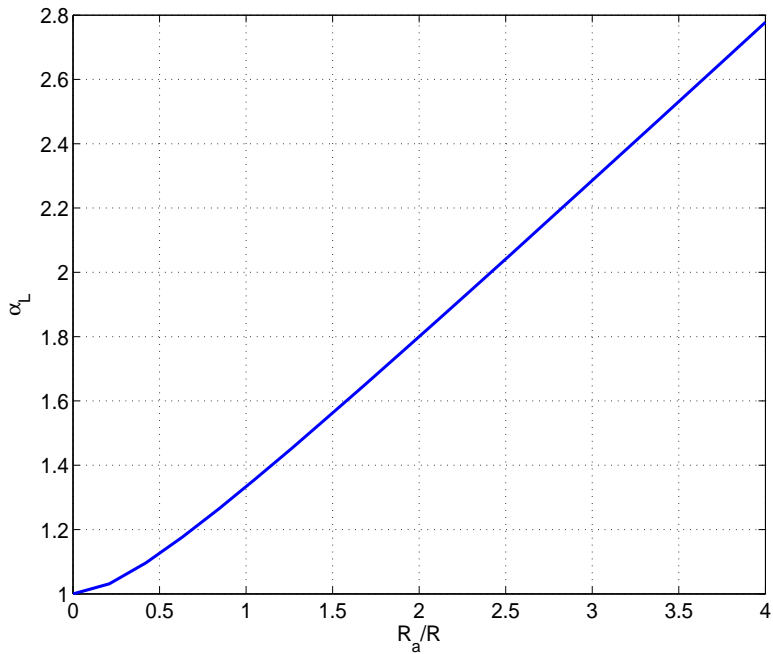}}
\caption{$\alpha_L$ versus $R_a/R$, showing that $\alpha_L$ grows sub-linearly with $R_a/R$.}
\label{fig:alphaL}
\end{figure}

Fig. \ref{fig:alphaL} depicts the variation of $\alpha_L$ over $q=R_a/R$. As seen, $\alpha_L$ grows sub-linearly with $R_a/R$. It is important to note that given $R$ and $R_a$, $\alpha_L$ and $\alpha_U$ will not be equal. In other words, unlike the upper bound, our proposed lower bound is not tight asymptotically.
The following lemmas help us prove Theorem \ref{th:alphaL}. While Lemma \ref{lem:ratio} holds for the asymptotic case, Lemma \ref{lem:firststepasymp} below is the asymptotic version of Lemma \ref{lem:firststep}. Also, Lemma \ref{lem:finalasymp} below can be interpreted as the asymptotic version of Lemma \ref{lem:Pr}. The proofs can be found in the appendix.

\begin{lem}\label{lem:firststepasymp}
Consider a dish attack of radius $R_a$ applied to $G(\lambda, R)$. Let $q=R_a/R$. If $\alpha<1+ \frac{q^2}{1+2q}$, we have
\begin{itemize}
\item $\bar{a}_1(\alpha-1)<\bar{a}$,

\item $\bar{a}_2(\alpha-1)<\bar{a}_1+\bar{a}$.
\end{itemize}
\end{lem} 

\begin{lem}\label{lem:finalasymp}
Consider the setting of Lemma \ref{lem:firststepasymp}. If $\alpha-1< \frac{q^2}{1+2q}$, for every $i\geq 1$ we have
\begin{align}
\bar{a}+\bar{a}_1+...+\bar{a}_i > \bar{a}_{i+1}(\alpha-1).
\end{align}
\end{lem} 

Now we are ready to prove Theorem \ref{th:alphaL}.

\emph{Proof of Theorem \ref{th:alphaL}:}
The proof is mostly along the same lines as for Lemma \ref{lem:a0fails} with a few minor changes. First note that as $\lambda$ grows very large, the network become connected. So there is no need to consider the subspace $G_c(\lambda, R)$ here. Second, $a$ is given as the initial number of failed nodes due to the attack. However, when $\lambda \rightarrow \infty$, $a$ tends to $\bar{a}$. Similarly, $a_i$ tends to $\bar{a}_i$ for $i\geq 1$.  Just like Lemma \ref{lem:a0fails}, in the best scenario, $\bar{a}_1(\alpha-1)$ is the excess capacity available to absorb the load from the $\bar{a}$ failed nodes.
If $\alpha-1< q^2/(1+2q)$, then we have $\bar{a}> \bar{a}_1(\alpha-1)$. In this case, $A_1$ cannot absorb the load of $A$, and the aggregate load of $\bar{a}+\bar{a}_1$ needs to be absorbed by the rest of the nodes. However, Lemma \ref{lem:finalasymp} asserts that such an absorbtion will not be realized as the failure propagates through $\mathcal{A}_2$, $\mathcal{A}_3$, and the outer rings until it takes out the whole network. \qedb

\section{Conclusion}\label{sec:conclusion}
This paper investigates the problem of cascading failures in finite-size networks modeled by random geometric graphs. Rigorous analytical results have been provided for studying the network resiliency under a dish attack of a given size. In particular, the average failure ratio due to the attack was studied in terms of the tolerance parameter, which is a critical design consideration in real-life networks. By deriving the lower and upper bounds on the average failure ratio, we were able to track the network reaction to different attacks. The asymptotic analysis of both bounds has also been presented.
Particularly, the asymptotic analysis of the upper bound revealed the threshold behavior of the network reaction to the changes in the tolerance parameter.
Our findings can be exploited to choose appropriate values of the tolerance parameter to avoid a cascade in a given network.

\appendix
\emph{Proof of Lemma \ref{lem:gaus}:}
First note that for $\bar{a}_1$ we have
\begin{align}\label{eq:lem1}
\notag \bar{a}_1=&\pi \lambda [(R_a+R)^2-R_a^2]= \pi \lambda R^2+ 2\pi \lambda  R R_a \\
&\geq 6+2 \sqrt{\pi^2 \lambda^2 R^2 R_a^2} \geq 6+2\sqrt{3\times 6}=6(1+\sqrt{2})>14.
\end{align}
For $i\geq 2$, $\bar{a}_i$ is equal to $\lambda$ times the area of $\mathcal{A}_i$.
Since the area of $\mathcal{A}_{i}$ is clearly larger than that of $\mathcal{A}_{i-1}$, we obtain $\bar{a}_{i}\geq \bar{a}_{i-1}$, which along with (\ref{eq:lem1}) leads to $\bar{a}_i\geq \bar{a}_1>14$ for $i\geq2$. \qedb

\emph{Proof of Lemma \ref{lem:dist}:}
Consider the angle $\hat{\theta}$ in Fig. \ref{fig:locationdist} and the arc of radius $r$ associated with $\hat{\theta}$ crossing over the node $i$. Let us denote the length of this arc by $\omega$.
The PDF of $r$ can be obtained by considering the probability of node $i$ being located inside the tiny area between the dashed lines,
thus we have
\begin{align}
\psi(r)=\frac{2\omega}{I(0,R_a, r_v, R)}.
\end{align}
Given $\hat{\theta}$, $\omega$ can be found as $\omega=\hat{\theta} \times r$. Now we only need to find $\hat{\theta}$. Note that $\hat{\theta}$ is an angle in a triangle with sides $r$, $r_v$, and $R$. Particularly, $\hat{\theta}$ is opposite to the side of length $R$. Therefore, we have
\begin{align}
\hat{\theta}=\arccos{\big(\frac{r_v^2-R^2+r^2}{2r_vr}\big)}.
\end{align}
The equation above holds when $r+r_v > R$.
For the case $r+r_v\leq R$, we simply have $\hat{\theta}=\pi$.
\qedb

\emph{Proof of Lemma \ref{lem:poi}:}
We first have
\begin{align}
&E[1/d_u]=\sum_{k=1}^{\infty} \frac{1}{k}\times \textrm{Pr}\{d_u=k \mid d_u\geq 1\}\\
&E[1/d_u^2]=\sum_{k=1}^{\infty} \frac{1}{k^2}\times \textrm{Pr}\{d_u=k \mid d_u\geq 1\}.
\end{align}
Since $d_u\sim \textrm{Poi}(\lambda_u)$, we have
\begin{align}
\textrm{Pr}\{d_u=k \mid d_u\geq 1\}=\frac{e^{-\lambda_u} \lambda_u^k}{k! (1-e^{-\lambda_u})}.
\end{align}
Therefore,
\begin{align}
&E[1/d_u]=\frac{e^{-\lambda_u}}{ 1-e^{-\lambda_u}} \sum_{k=1}^{\infty} \frac{1}{k} \frac{\lambda_u^k}{k!} \label{eq:sum1}\\
&E[1/d_u^2]=\frac{e^{-\lambda_u}}{ 1-e^{-\lambda_u}} \sum_{k=1}^{\infty} \frac{1}{k^2} \frac{\lambda_u^k}{k!}. \label{eq:sum2}
\end{align}
To find a closed-form for the summations above, we have
\begin{align}\label{eq:S}
\sum_{k=0}^{\infty} \frac{1}{k+1} \frac{\lambda_u^{k+1}}{(k+1)!}=
\int \sum_{k=0}^{\infty} \frac{\lambda_u^k}{(k+1)!} \quad d\lambda_u.
\end{align}
For the expression under the integral, we could use the Taylor expansion of an exponential function as
\begin{align}\label{eq:underinteg}
\sum_{k=0}^{\infty} \frac{\lambda_u^k}{(k+1)!}=\frac{1}{\lambda_u} \sum_{k=1}^{\infty} \frac{\lambda_u^k}{k!}= \frac{e^{\lambda_u}-1}{\lambda_u}.
\end{align}
Substituting (\ref{eq:underinteg}) into (\ref{eq:S}) yields (\ref{eq:g}). Note that the integral in (\ref{eq:g}) could be evaluated numerically. Along the same lines, we could find the following for the summation in (\ref{eq:sum2}):
\begin{align}\label{eq:sum2closed}
\notag \sum_{k=1}^{\infty} \frac{1}{k^2} \frac{\lambda_u^k}{k!}=&\int_{-\infty}^{\lambda_u} \sum_{k=1}^{\infty} \frac{1}{k} \frac{x^{k-1}}{k!} dx= \\
&\int_{-\infty}^{\lambda_u} 1/x \int_{-\infty}^{x} \sum_{k=0}^{\infty} \frac{y^k}{(k+1)!} dy=\int_{-\infty}^{\lambda_u} \frac{g(x)}{x} dx.
\end{align}
Substituting (\ref{eq:sum2closed}) in (\ref{eq:sum2}) gives us (\ref{eq:poi2}). \qedb

 \emph{Proof of Lemma \ref{lem:i}:}
The load of $i$ will be redistributed equally among its neighbors outside $\mathcal{A}$. Let us denote the number of such neighbors by $d_u$. The average load received by $v$ is $h^{(1)}(r)=E[1/d_u | r, d_u>0]$ where the average is taken over $d_u$.
Given that $i$ is located at $r$, $d_u$ is distributed as  $\textrm{Poi}\big(\lambda_u=\lambda (\pi R^2-I(r,R, 0,R_a))\big)$. As a result, by applying Lemma \ref{lem:poi} we obtain
\begin{align}
h^{(1)}(r)= E[1/d_u | r, d_u>0]=\frac{e^{-\lambda J(r) }g(\lambda J(r))}{1-e^{-\lambda J(r)}}.
\end{align}
If $i$ is located randomly  and uniformly on $\mathcal{I}(r_v,R,0,R_a)$, the probability density function of its distance from the center is given by $\psi(r)$ at (\ref{eq:phi}). Applying $\psi(r)$ to find $E[h(r)^{(1)}]$ and $E[h(r)^{(2)}]$ by taking average over $r$ leads to (\ref{eq:aveli}) and (\ref{eq:aveli2}).
When $R-r_v\geq R_a$, the attacked region entirely resides in $v$'s neighborhood; hence, the averaging of (\ref{eq:aveli}) or (\ref{eq:aveli2}) over $r$ is no longer needed, and we obtain (\ref{eq:lihalf}). \qedb

\emph{Proof of Theorem \ref{thm:aveload}:}
Let us denote the number of neighbors of $v$ inside $\mathcal{I}(r_v,R,0,R_a)$ by $N$. Note that $N\sim\textrm{Poi}\big(\lambda I(r_v,R,0,R_a) \big)$. Let $l_u, u=1,...,N$, denote the sequence of r.v.'s corresponding to the load redistributed to $v$ by its neighbors. As it has been shown in the proof of Lemma \ref{lem:i}, $l_u$ completely depends on a Poisson point process outside $\mathcal{I}(r_v,R,0,R_a)$, whereas $N$ is given by a Poisson point process inside $\mathcal{I}(r_v,R,0,R_a)$. Hence, $N$ and the random variables $l_u, u=1,...,N$ are independent. Therefore, we can write
\begin{align}
E[L_v]&= E[\sum_{u=1}^N l_u]=E\big[E[\sum_{u=1}^N l_u | N=n]\big]\\
&\overset {(a)}{=} E[N]\times E[l_u] \overset{(b)}{=} \lambda I(r_v,R,0,R_a) \times  \int_{r_v-R}^{R_a} h(r) \times \psi(r) \quad dr,
\end{align}
where $(a)$ follows from Wald's identity \cite{RossProbability} and the fact that a Poisson point process, given the number of points, becomes a uniform point process. Also, $(b)$ follows from replacing $E[l_u]$ with (\ref{eq:aveli}) from Lemma \ref{lem:i}. Similarly, for $\sigma_L$ we have
\begin{align}
\sigma_{L_v}^2&= E[(L_v-\bar{L}_v)^2]=E\big[E[(\sum_{u=1}^N (l_u-\bar{l}_u))^2 | N=n]\big]\\
&=E[N]\times E[(l_u-\bar{l}_u)^2]= \lambda I(r_v,R, 0,R_a)\times \sigma_{l_u}^2.
\end{align}
When $R-r_v\geq R_a$, $\mathcal{A}$ is entirely included in $v$'s neighborhood. Thus, $E[N]=\lambda\pi R_a^2$, and $E[l_u]$ is given by (\ref{eq:lihalf}), together leading to (\ref{eq:Lhalf}). \qedb

\emph{Proof of Theorem \ref{th:loadprob}:}
Since $v$ is located randomly and uniformly in $\mathcal{A}_1$, the PDF of $r_v$ is obtained as $\frac{2\pi r_v}{|\mathcal{A}_1|}$. Thus we have
\begin{align}
p_1=\int_{R_a}^{R_a+R} \textrm{Pr}\{L_{v}\leq \alpha-1\} \times \frac{2\pi r_v}{|\mathcal{A}_1|}\ dr_v.
\end{align}
It remains to find $\textrm{Pr}\{L_{v}\leq \alpha-1\}$. According to Corollary \ref{cor normala:pprox}, if $\lambda\times I(r_v,R, 0,R_a)>>1$, we have $\textrm{Pr}\{L_{v}\leq \alpha-1\}\approx \Phi\big(\frac{\alpha-1-\bar{L}_v}{\sigma_{L_v}}\big)$. However, this holds either when $r_v$ is close to $R_a$ or when $\lambda$ is large. If $v$ is located far from the edge of the attack, $I(r_v,R, 0,R_a)$ might be small, and $\lambda\times I(r_v,R, 0,R_a)>>1$ may not hold for moderate values of $\lambda$. However, as $I(r_v,R, 0,R_a)$ becomes smaller, $J(\cdot)$ in (\ref{eq:lgivenr}) and (\ref{eq:l2givenr}) becomes larger for the neighbors of $v$ in $\mathcal{A}$, causing $E[l_u]$ and $E[l_u^2]$ to drop quickly for these neighbors. At the same time, the number of such neighbors, which is a Poisson r.v. with mean $\lambda\times I(r_v,R, 0,R_a)$, becomes smaller. Therefore, as $r_v$ grows larger, both $\bar{L}_v$ and $\sigma_{L_v}$ in (\ref{eq:aveload}) drop quickly until they become zero when $r_v=R_a+R$. As a result, $\textrm{Pr}\{L_{v}\leq \alpha-1\}$ grows rapidly as $r_v$ increases, and becomes very close to 1. This can also be verified using numerical methods.
Now, since $\Phi\big(\frac{\alpha-1-\bar{L}_v}{\sigma_{L_v}}\big)$ also takes values very close to one in such cases, approximating $\textrm{Pr}\{L_{v}\leq \alpha-1\}$ by $\Phi\big(\frac{\alpha-1-\bar{L}_v}{\sigma_{L_v}}\big)$ will have a negligible effect on the value of $p_1$. Applying such an approximation leads to (\ref{eq:loadprob}). \qedb

\emph{Proof of Theorem \ref{th:upper}:}
Since $0\leq f\leq 1$, we have
\begin{align}\label{eq:fproof}
\notag \bar{f}\leq \textrm{Pr}\{f=0\}\times 0+ \textrm{Pr}\{f>0\}\times 1= \textrm{Pr}\{f>0\}&=\textrm{Pr}\{\textrm{at least 1 failure in } \mathcal{S}\setminus \mathcal{A}\}\\
&=\textrm{Pr}\{\textrm{at least 1 failure in } \mathcal{A}_1\}=1-p_0,
\end{align}
where $p_0\triangleq$ Pr$\{$no failures in $\mathcal{A}_1\}$. Note that $p_0$ is the probability that the load  received by every node in $\mathcal{A}_1$ is less than $\alpha-1$. Let us denote by $P(k)$ the probability that there are $k$ nodes in $\mathcal{A}_1$. Also recall that $p_1$ is the probability that the load received by a node located randomly and uniformly in $\mathcal{A}_1$ is less than $\alpha-1$. We then have
\begin{align}\label{eq:p0}
p_0=\sum_{k=0}^{\infty} P(k) p_1^k=\sum_{k=0}^{\infty} \frac{e^{-\lambda_1}}{k!} \lambda_1^k p_1^k=e^{-\lambda_1} \sum_{k=0}^{\infty} \frac{(\lambda_1 p_1)^k}{k!}=e^{-\lambda_1} e^{\lambda_1 p_1}=e^{-\lambda_1(1-p_1)}.
\end{align}
Substituting $p_1$ above by its value given by Theorem \ref{th:loadprob}, and then substituting (\ref{eq:p0}) into (\ref{eq:fproof}), we obtain (\ref{eq:f}). \qedb

\emph{Proof of Theorem \ref{th:Lasymp}:}
In order to prove the theorem, we will show that, as $\lambda\rightarrow \infty$, we have $\sigma_{L_v}\rightarrow 0$ , and $E[L(r_v)]$ takes the righthand side of (\ref{eq:Lasymp}). First note that when $\lambda \rightarrow \infty$, by applying the L'Hopital's rule \cite{RossProbability} to (\ref{eq:lgivenr}) and (\ref{eq:l2givenr}) we obtain
\begin{align}
&h^{(1)}(r)=E[l_u| r]\underset  {\lambda\rightarrow \infty} {\rightarrow} \frac{1} {\lambda J(r)},\label{eq:h1asymp}\\
&h^{(2)}(r)=E[l_u^2| r] \underset  {\lambda\rightarrow \infty} {\rightarrow} \frac{1} {[\lambda J(r)]^2}\label{eq:h2asymp}.
\end{align}
Substituting (\ref{eq:h1asymp}) into (\ref{eq:aveload}) gives us the asymptotic average of $L(r_v)$ in (\ref{eq:Lasymp}). Now let us show that the asymptotic value of $\sigma_{L_v}$ tends to 0. Using (\ref{eq:aveli}), (\ref{eq:aveli2}), and (\ref{eq:sigmali}), we find that
\begin{align}
\underset  {\lambda\rightarrow \infty} {\sigma_{l_u}^2} \rightarrow \frac{m(r_v)}{\lambda^2},
\end{align}
where
\begin{align}
\notag m(r_v)= \frac{2}{I(r_v, R, 0, R_a)} \int_{r_v-R}^{R_a} \frac{ r } {[J(r)]^2} &\arccos{\big(\frac{r_v^2-R^2+r^2}{2r_vr}\big)} dr \\
&- \frac{4}{[I(r_v, R, 0, R_a)]^2} \bigg[\int_{r_v-R}^{R_a} \frac{ r } {J(r)} \arccos{\big(\frac{r_v^2-R^2+r^2}{2r_vr}\big)} dr\bigg]^2
\end{align}
is a function of $r_v$, taking only finite values. Now considering (\ref{eq:sigmaload}) for $\sigma_{L_v}^2$ we have
\begin{align}
\sigma_{L_v}^2=\lambda I(r_v,R, 0,R_a)\times \sigma_{l_u}^2 \underset  {\lambda\rightarrow \infty} {\rightarrow} \frac{I(r_v,R, 0,R_a)}{\lambda} m(r_v)\underset  {\lambda\rightarrow \infty}{\rightarrow} 0.
\end{align} \qedb

\emph{Proof of Theorem \ref{th:cascade}:}
In the best case, let us assume that all the nodes located at $r>r_v$ are healthy and have received no load so far. We will show that $v$ still fails in this case. Note that this assumption is equivalent to having a dish attack of radius $r_v\geq R_a$. Let us denote by $L(r)$ and $L^{\prime}(r)$ the asymptotic load distribution right after the attack for attacks of radius $R_a$ and $r_v$, respectively. Since the latter is a larger attack, we have
\begin{align}
L^{\prime}(r_v)\geq L(R_a)=\alpha_U-1>\alpha-1.
\end{align}
Therefore node $v$ fails. Applying the same procedure to nodes located at $r>r_v$ results in a propagation of failures throughout the network, leading to $\bar{f}=1$. \qedb

\emph{Proof of Theorem \ref{th:upperasymp}:}
Looking at (\ref{eq:Lasymp}) in Theorem \ref{th:Lasymp}, it could be concluded that in the asymptotic case the closer a node is to the attack region (i.e., the smaller is $r_v$), the larger is the load it receives right after the attack.
Therefore, we have $L(R_a)>L(r_v)$, for $r_v>R_a$. Consequently, for $\alpha\geq\alpha_U$, we have
\begin{align}
L(r_v)<L(R_a)= \alpha_U-1\leq \alpha-1,
\end{align}
which means that none of the nodes in $\mathcal{A}_1$ would fail. Hence, there will not be any propagation of failures, resulting in $\bar{f}=0$.
It remains to prove that a cascade of failures is assured when we have $\alpha<\alpha_U$. In this case we have $L(R_a)=\alpha_U-1>\alpha-1$, which means that every node located at $R_a$ would fail. Now, by simply applying Theorem \ref{th:cascade} for $r_v=R_a$, we have $\bar{f}=1$. \qedb

\emph{Proof of Lemma \ref{lem:firststep}:
}We need to show that
\begin{align}
\notag &\frac{\bar{a}_2}{\bar{a}_1} =\frac{(R_a+2R)^2-(R_a+R)^2}{(R_a+R)^2- R_a^2}=1+\frac{2R}{R+2R_a}< 1+\frac{1}{\alpha-1},\\
\notag &\Rightarrow \alpha-1< \frac{R+2R_a}{2R} =1/2 + \frac{R_a}{R}\\
&\Rightarrow \alpha-1<1/2 + q. \label{eq:2asymp}
\end{align}
However, (\ref{eq:2asymp}) is given by the lemma's assumption, which completes the proof. \qedb

\emph{Proof of Lemma \ref{lem:ratio}:}
We need to show that
\begin{align}\label{eq:ratio1}
\notag &\bar{a}_{i+1}\times \bar{a}_{i-1} \leq \bar{a}_i^2\\
\notag &\Rightarrow (R_{i+1}^2-R_i^2) (R_{i-1}^2-R_{i-2}^2) \leq (R_i^2 - R_{i-1}^2)^2 \\
& \Rightarrow \big(2R_a + (2i+1)R\big) \big(2R_a +(2i-3)R\big) \leq \big(2R_a+(2i-1)R\big)^2.
\end{align}
If we set $x_1=2R_a + (2i+1)R$ and $x_2=2R_a +(2i-3)R$, (\ref{eq:ratio1}) could be deducted from the ``inequality of arithmetic and geometric means" \cite{Steele04}, asserting that for two non-negative numbers $x_1$ and $x_2$ we have
\begin{align}
\frac{x_1+x_2}{2}\geq \sqrt{x_1x_2}  \Rightarrow (\frac{x_1+x_2}{2})^2\geq x_1x_2.
\end{align}
Now that (\ref{eq:ratiofirst}) is proved, (\ref{eq:ratiosec}) could be obtained  by simply applying the second item of Lemma \ref{lem:firststep}. \qedb

\emph{Proof of Lemma \ref{lem:Pr}:} We prove the theorem by induction. For $i=0$ the equality holds. Let us assume that the theorem holds for $i=k-1, k>1$; we prove that it holds for $i=k$ as well.
We have
\begin{align}
\textrm{Pr}\{a_0+a_1+...+a_k > a_{k+1}(\alpha-1)\}\overset {(a)}{\geq} & \textrm{Pr}\{a_0+a_1+...+a_{k-1}+ a_k > a_k \alpha\}\\
&=\textrm{Pr}\{a_0+a_1+...+a_{k-1}> a_k(\alpha-1)\}\overset{(b)}{\geq} \textrm{Pr}\{a_0> a_1(\alpha-1)\},
\end{align}
where (a) holds by applying Lemma \ref{lem:ratio} and the Gaussian distribution of $a_i, i\geq1$. In the above, (b) holds due to the induction assumption made for $i=k-1$. \qedb

\emph{Proof of Lemma \ref{lem:a0fails}:}
Here $a_0$ is given as the initial number of failed nodes due to the attack.
At its best, $a_1(\alpha-1)$ is the excess capacity available to absorb the load of these $a_0$ failed nodes. Now consider a best-case load distribution strategy where all the nodes in $\mathcal{A}$ can collaborate and all the nodes in $\mathcal{A}_1$ are connected to $\mathcal{A}$. Then, nodes in $\mathcal{A}$ can distribute their loads equally among the nodes in set $A_1$ in order to use all the excess capacity and avoid a cascade. If $a_0> a_1(\alpha-1)$, $A_1$ cannot absorb the load of $A$, and the aggregate load of $a_0+a_1$ needs to be absorbed by rest of the nodes. However, Lemma \ref{lem:Pr} asserts that such an absorbtion becomes even less likely, and the failure propagates through $\mathcal{A}_2$, $\mathcal{A}_3$, and outer rings until it takes out the whole network. Therefore, the probability of a total failure is lower-bounded by $\textrm{Pr}\{a_0> a_1(\alpha-1)\}$. \qedb

\emph{Proof of Lemma \ref{lem:firststepasymp}:}
In order to prove the first item, note that we have
\begin{align}\label{eq:1}
\bar{a}_1(\alpha-1)<\bar{a} \Rightarrow \alpha-1<\frac{\bar{a}}{\bar{a}_1}= \frac{R_a^2}{R^2+2R_aR}= \frac{q^2}{1+2q}.
\end{align}
Moving on to prove the second item, given that $a_1(\alpha-1)<a_0$, it suffices to show that
\begin{align}
\bar{a}_2(\alpha-1)< \alpha \bar{a}_1= \bar{a}_1+\bar{a}_1(\alpha-1) <\bar{a}_1+\bar{a}.
\end{align}
From Lemma \ref{lem:firststep}, we already know
\begin{align}\label{eq:2}
\alpha-1< 1/2 + q \Rightarrow \bar{a}_1\times \alpha> \bar{a}_2 (\alpha-1).
\end{align}
To complete the proof, we only need to show that (\ref{eq:1}) leads to (\ref{eq:2}), i.e., we need to have
\begin{align}\label{eq:3}
\frac{q^2}{1+2q} < 1/2+ q.
\end{align}
Inequality (\ref{eq:3}) can be rewritten as
\begin{align}\label{eq:4}
q^2+2q+1/2>0 \Rightarrow (q+1)^2-1/2 >0 \Rightarrow (q+1-\frac{1}{\sqrt{2}})(q+1+\frac{1}{\sqrt{2}}) >0,
\end{align}
which always holds since $q>0$. This completes the proof. \qedb

\emph{Proof of Lemma \ref{lem:finalasymp}:}
The proof is simple and is given by induction on $i$. First, note that for $i=1$, the case is proven by Lemma \ref{lem:firststepasymp}. Now suppose that the statement is true for $i=k-1$. We show that it holds for $i=k$, $k\geq2$, as well. For $i=k$ we have
\begin{align}
\notag (a_0+\bar{a}_1+...+\bar{a}_{k-1}) +\bar{a}_k > &\bar{a}_k(\alpha-1) + \bar{a}_k\\
&= \bar{a}_k \times\alpha \geq \bar{a}_{k+1}(\alpha-1),
\end{align}
where the last inequality holds due to Lemma \ref{lem:ratio}. \qedb

\newpage
\bibliographystyle{ieeetran}
\bibliography{biblio1}

\end{document}